# PRE-MAIN SEQUENCE ACCRETION IN THE LOW METALLICITY GALACTIC STAR-FORMING REGION Sh 2-284

V. M. Kalari & J. S. Vink
Armagh Observatory, College Hill, Armagh, BT61 9DG, UK; vek@arm.ac.uk

## ABSTRACT

We present optical spectra of pre-main sequence (PMS) candidates around the H$\alpha$ region taken with the Southern African Large Telescope, SALT, in the low metallicity ($Z$) Galactic region Sh 2-284, which includes the open cluster Dolidze 25 with an atypical low metallicity of $Z \sim 1/5\ Z_{\odot}$. It has been suggested on the basis of both theory and observations that PMS mass-accretion rates, $\dot{M}_{\rm acc}$, are a function of $Z$. We present the first sample of spectroscopic estimates of mass-accretion rates for PMS stars in any low-$Z$ star-forming region. Our data-set was enlarged with literature data of H$\alpha$ emission in intermediate-resolution $R$-band spectroscopy. Our total sample includes 24 objects spanning a mass range between 1 - 2 $M_{\odot}$ and with a median age of approximately 3.5 Myr. The vast majority (21 out of 24) show evidence for a circumstellar disk on the basis of 2MASS and *Spitzer* infrared photometry. We find $\dot{M}_{\rm acc}$ in the 1 - 2 $M_{\odot}$ interval to depend quasi-quadratically on stellar mass, with $\dot{M}_{\rm acc} \propto M_{*}^{2.4\,\pm\,0.35}$, and inversely with stellar age $\dot{M}_{\rm acc} \propto t_{*}^{-\,0.7\,\pm\,0.4}$. Furthermore, we compare our spectroscopic $\dot{M}_{\rm acc}$ measurements with solar $Z$ Galactic PMS stars in the same mass range, but, *surprisingly find no evidence for a systematic change in $\dot{M}_{\rm acc}$ with $Z$*. We show that literature accretion-rate studies are influenced by detection limits, and we suggest that $\dot{M}_{\rm acc}$ may be controlled by factors other than $Z_{*}$, $M_{*}$, and age.

*Subject headings:* stars: pre main-sequence, stars:variables: T Tauri, Herbig Ae/Be, accretion disks, stars: formation, stars: fundamental parameters

## 1. INTRODUCTION

Present understanding suggests that most stars accrete mass from a circumstellar disk during their pre-main sequence (PMS) phase (see Hartmann 2008). The rate of mass accretion ($\dot{M}_{\rm acc}$) from the disk to the star is vital to describe the system's evolution, including the potential growth of planets in the disk whilst the star reaches its main sequence configuration. Viscous disk evolution models predict that $\dot{M}_{\rm acc}$ drops with time as $\dot{M}_{\rm acc} \propto t_{*}{}^{-1.5}$, a prediction which has been substantiated by empirical studies (Hartmann et al. 1998). Observations have also found that the mass-accretion rate scales with stellar mass as $\dot{M}_{\rm acc} \propto M_{*}{}^{\alpha}$ over a mass range from 0.3 $M_{\odot}$-3 $M_{\odot}$, with index $\alpha \sim 2$ (Calvet et al. 2004; Natta et al. 2006; Muzerolle et al. 2005; Sicilia-Aguilar et al. 2006; Manara et al. 2012), though it has been found to be discrepant at $M_{*} < 0.8\,M_{\odot}$ (Fang et al. 2009; Barentsen et al. 2011). Note that the $\dot{M}_{\rm acc}$ vs. Mass relation is not reproduced by the current paradigm of disk evolution, and it is subject to ongoing research (e.g. Hartmann et al. 2006).

It has also been suggested that $\dot{M}_{\rm acc}$ is a function of metallicity, $Z$. Seven PMS star candidates discovered in the Large Magellanic Cloud (LMC) at $Z \approx 0.006$ by Beaulieu et al. (1996) were the first PMS candidates identified in a low-$Z$ region. Beaulieu et al. (1996) selected these objects because they exhibited peculiar photometric variability similar to Galactic Herbig HAe/Be stars. The number of low-$Z$ PMS candidates was increased in the follow-up works of Lamers et al. (1999); de Wit et al. (2002); de Wit et al. (2003) and de Wit et al. (2005) using similar techniques, where the authors noted that these stars also exhibited H$\alpha$ emission and were located in H II regions. These studies showed that the luminosities of these objects were significantly higher than those of Galactic PMS stars at solar $Z_{\odot}$, and with similar spectral types. de Wit et al. (2003) showed that HAeBe candidates in the Small Magellanic cloud (SMC) at lower metallicities ($Z \sim 0.002$ - $0.004$) were even brighter than their LMC counterparts. Based on the inferred luminosities of the Magellanic clouds (MCs) PMS stars, de Wit et al. (2005) suggested that the protostellar mass-accretion rates are inversely proportional to metallicity, leading to more luminous PMS stars that are visible at younger ages at lower metallicities. Whether the PMS $\dot{M}_{\rm acc}$ are analogously related to metallicity is not clear. The $\dot{M}_{\rm acc}$ estimated by de Wit et al. (2005) for their best LMC PMS candidate (ELHC 7) indicated no significant difference in comparison to the canonical Galactic average.

On the contrary, studies based on the Balmer continuum excess (Romaniello et al. 2004) and H$\alpha$ emission (Panagia et al. 2000) inferred higher $\dot{M}_{\rm acc}$ compared to the Galactic average in a LMC region close to SN1987A. Follow-up H$\alpha$ *photometric* studies by the same group in the LMC regions around SN1987A (De Marchi et al. 2010) and 30 Doradus (De Marchi et al. 2011a) which were summarized in Spezzi et al. (2012); and in the SMC open clusters NGC 346 ($Z \sim 0.002$; De Marchi et al. 2011b) and NGC 602 ($Z \sim 0.004$; De Marchi et al. 2013) reported a metallicity dependence on $\dot{M}_{\rm acc}$ (see below) whilst also finding $\dot{M}_{\rm acc} \propto M_{*}{}^{1}$, and $\dot{M}_{\rm acc} \propto t_{*}{}^{0.36}$ in the 1-2 $M_{\odot}$ range for the LMC PMS stars. They em-ployed a novel method using $VI$H$\alpha$ photometry, rather than conventional spectroscopy to derive $\dot{M}_{\rm acc}$. We have applied a similar method in the Galactic regions IC 1396 (Barentsen et al. 2011), NGC 2264 (Barentsen et al. 2013) and NGC 6530 (Kalari et al. 2014, in prep.), and found that the results from the photometric method do not differ significantly from spectroscopic H$\alpha$ or $U$-band measurements, nor did we find a large number of interlopers. Furthermore, we discovered an accreting PMS candidate star in the LMC region 30 Doradus (Kalari et al. 2014), which is within $\sim 10$ pc to one of the regions examined in Spezzi et al. (2012). These various results suggest there may well be genuine highly-accreting PMS stars present in the MCs. De Marchi et al. (2013) suggest accretion rates are inversely proportional to metallicity based on studies of the two SMC open clusters. De Marchi et al. (2011b) reported a median $\dot{M}_{\rm acc}$ of $4\times10^{-8}\,M_{\odot}yr^{-1}$ in NGC 346 for a bimodal age distribution of 1 and 20 Myr; and De Marchi et al. (2013) found a median of $2.5\times10^{-8}\,M_{\odot}yr^{-1}$ for a 2 Myr population in NGC 602 and $4\times10^{-9}\,M_{\odot}yr^{-1}$ for a 20 Myr population. De Marchi et al. (2013) summarized their work in the MCs using an equation of the form $\log M_{\rm acc} = a\times\log t_{*} + b\times\log M_{*} + c$, where $a, b$ and $c$ are constants, and $c$ is inversely proportional to Z. Similarly, Spezzi et al. (2012) suggest that $\dot{M}_{\rm acc}$ is inversely proportional to $Z$, and suggest that $\dot{M}_{\rm acc}$ in the LMC is higher than that of Galactic PMS stars having similar masses, irrespective of age. Spezzi et al. (2012) found a median $\dot{M}_{\rm acc}\sim5\times10^{-8}\,M_{\odot}yr^{-1}$ of LMC PMS stars having an average age of 10 Myr, when accretion in most low-mass ($<0.8\,M_{\odot}$) Galactic PMS stars is thought to have ceased (Fedele et al. 2010). Approximately, for 1 - 2 $M_{\odot}$ the PMS lifetime is between 30 - 5 Myr respectively (Palla & Stahler 1999). This suggests that accretion is a dominant process for a large fraction of the PMS lifetime in the LMC, contrary to current observations in the Galaxy. Prolonged accretion at these rates suggests accretion adds significant mass to the central star after the 'birthline' (Stahler 1983; Palla & Stahler 1990; Hartmann et al. 1997; Tout et al. 1999) – contrary to current disk or spherical accretion models. The authors suggest that the lower disk opacity, and viscosity due to lowering $Z$ leads to a longer viscous timescale, thereby allowing accretion to take place at such older ages.

However, observational studies of disk lifetime at lower $Z$ indicate that metal-poor stars in the extreme outer Galaxy ($Z\approx0.002$) lose their disks significantly more quickly than solar-$Z$ PMS stars (Yasui et al. 2009). Accretion at the high rates measured in the MCs by Spezzi et al.(2012) and De Marchi et al.(2010;2011a;2011b;2013) means that the disks of these PMS stars have undergone gravitational instability (Pringle 1981; Hartmann 2006; Cai et al. 2006), shortening the disk lifetime. Furthermore, lowering $Z$ lowers the dust shielding efficiency against UV photoevaporation from OB stars leading to possible disk erosion, an effect which has been demonstrated in the LMC (Spezzi et al. 2012). Internal photoevaporation is also thought to reduce disk lifetimes at low-$Z$, as the lower metal content increases X-ray opacity (Ercolano & Clarke 2010). In all these cases disk lifetimes in low-$Z$ environments are thought to be reduced, suggesting that it is challenging for accretion to proceed at significant rates at older ages in low $Z$ environments. It also suggests that $\dot{M}_{\rm acc}$ may be higher at lower $Z$, but only at young ages.

It is thus essential to further scrutinize the differences in $\dot{M}_{\rm acc}$ of metal-poor PMS stars in comparison to solar $Z$ PMS stars, as to understand star formation in different $Z$ environments, as well as the processes involved in disk evolution. To this purpose, we employ spectroscopic observations of candidate PMS stars located in the Galactic star-forming region Sh 2-284 (Sharpless 1959). Sh 2-284 is an H II region encompassing the central open cluster Dolidze 25 (also OCL-537). The B stars of Dolidze 25 have been shown to be metal deficient with respect to solar values, with all measured elemental abundances ranging from $-0.6$ to $-0.9$ dex, leading to an approximate metal deficiency of $-0.7$ dex (Lennon et al. 1990; Fitzsimmons et al. 1992). The association of Dolidze 25 with the Sh 2-284 indicates that the Sh 2-284 region is also characterized by low metallicity (Puga et al. 2009; Delgado et al. 2010; Cusano et al. 2011).

Sh 2-284 is located at a distance of approximately 4 kpc (Delgado et al. 2010; Cusano et al 2010) and provides a unique opportunity to observe PMS stars in the 1 - 2 $M_{\odot}$ range with medium-resolution spectroscopy, and we can compare our measured accretion rates with literature spectroscopic measurements of solar $Z$ PMS stars, as well as results claimed from H$\alpha$ photometry in the LMC by Spezzi et al. (2012) [1].

The paper is organized as follows. Section 2 details the data sample, and measured properties of the PMS stars. Section 3 describes our results and method to calculate $\dot{M}_{\rm acc}$. A discussion of the $\dot{M}_{\rm acc}$ distribution with respect to $M_{*}$, $t_{*}$ and $Z$ is presented in Section 4. The conclusions of our work are summarized in Section 5.

## 2. DATA

For our program, we selected PMS candidates on the basis of infrared 2MASS and *Spitzer* photometry. Six *Spitzer* identified Class II sources Young Stellar Objects (YSO) from Puga et al. (2009) with near-infrared *JHKs* magnitudes from 2MASS (Two Micron all sky survey; Cutri et al. 2003) or optical *UBV* magnitudes from Delgado et al. (2010) were selected. This information is presented in Table 1. The location of the selected stars is overlaid on an H$\alpha$ gray-scale in Fig. 1. Long-slit spectra of these objects were obtained using the Robert Stobie Spectrograph (RSS) on the 10 metre Southern African Large Telescope (SALT) located in Sutherland, South Africa (Burgh et al. 2003). A slit width of $1.25''$ was used to obtain good sky subtraction. A red grating covering $\lambda\lambda$ 5700 - 7500 Å with resolution $R\sim2000$, and dispersion 1 Å per pixel was used. Observations were obtained over 3 nights during December 2013. The seeing varied between 1.1 and $1.5''$.

Bias subtraction, flat-field correction, and cosmic ray removal were performed using standard IRAF[2] proce-dures. Wavelength calibration was performed using Ne arc spectra using the IRAF *identify* task. Spectra were corrected for sensitivity using the Sutherland extinction curve. The variable pupil design of the SALT telescope makes absolute flux calibration impossible. The final extracted 1-dimensional spectra are shown in Fig 2. The average signal to noise is $\sim$ 20 - 40.

[1] Due to the lack of publicly available data of accretion properties of PMS stars in the SMC, we restrict our comparison to the LMC PMS sample of Spezzi et al. (2012).

[2] IRAF is distributed by the National Optical Astronomy Observatory, which is operated by the Association of Universities for Research in Astronomy, Inc., under cooperative agreement with the National Science Foundation.

TABLE 1
LITERATURE DATA FOR OBSERVED STARS

| 2MASS | $\alpha$ (J2000) | $\delta$ (J2000) | $V$ | $B-V$ | $U-B$ | $J$ | $J-H$ | $J-Ks$ | D10[a] | P09[b] |
|---|---|---|---|---|---|---|---|---|---|---|
| 06443682+0016186 | 06 44 36.8 | +00 16 18.70 | ... | ... | ... | 14.58 | 0.74 | 1.54 | ... | 17 |
| 06445577+0013168 | 06 44 55.8 | +00 13 17.30 | 15.23 | 0.32 | 0.65 | 13.89 | 0.43 | 0.99 | 307 | 37 |
| 06445837+0014151 | 06 44 58.4 | +00 14 15.50 | 16.68 | 0.38 | 1.08 | 14.47 | 0.55 | 0.80 | 585 | 44 |
| 06450208+0019443 | 06 45 02.1 | +00 19 44.00 | ... | ... | ... | 14.62 | 0.49 | 0.78 | ... | 50 |
| 06450681+0013535 | 06 45 06.8 | +00 13 54.00 | 16.59 | 0.61 | 1.13 | 13.95 | 0.77 | 1.57 | 1563 | 52 |
| 06451007+0014117 | 06 45 10.1 | +00 14 12.10 | 15.71 | 0.58 | 0.91 | 13.37 | 0.00 | 1.61 | 1897 | 55 |

[a] Identification no. in Delgado et al. (2010).
[b] Identification no. in Puga et al. (2009).

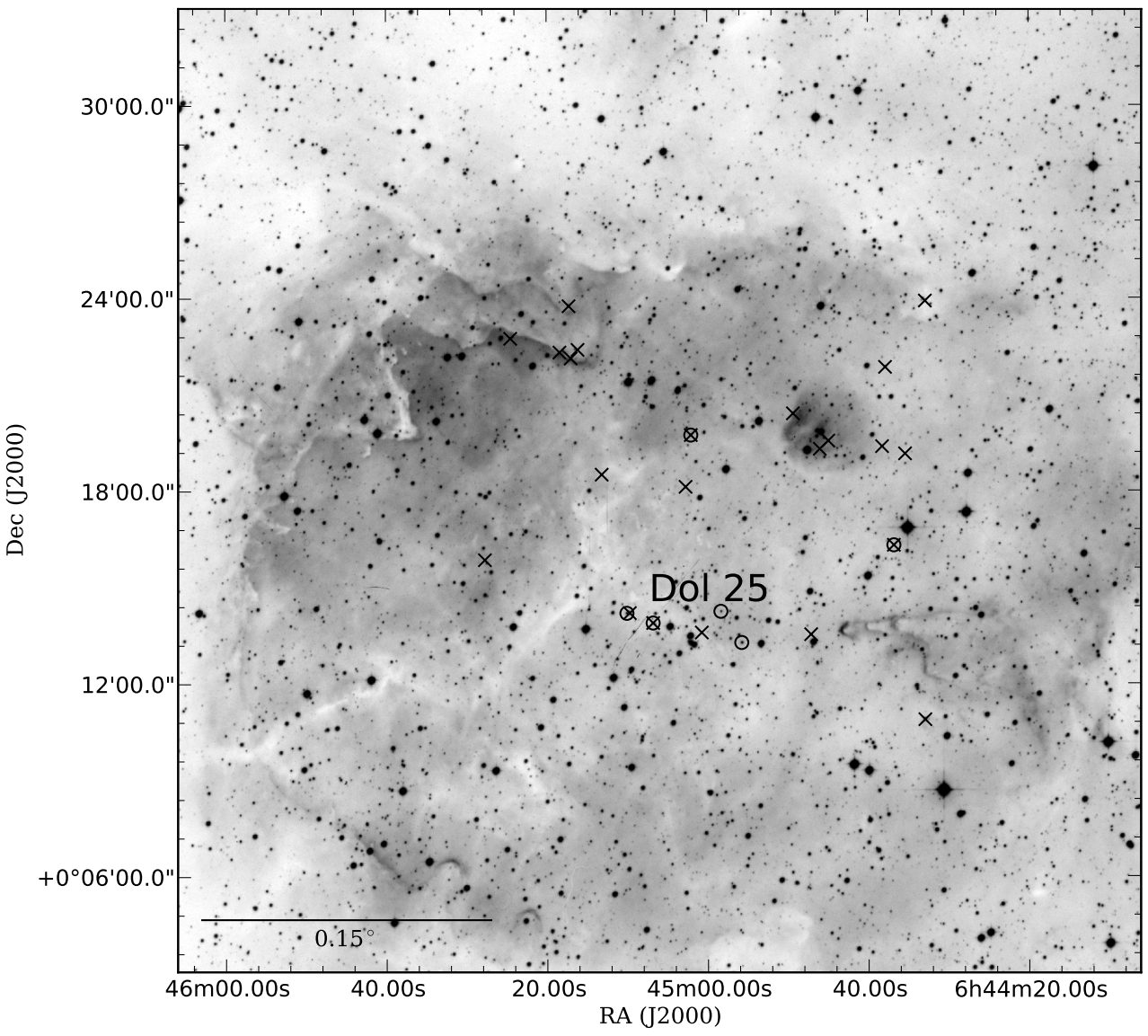

FIG. 1.— Sh 2-284 in H$\alpha$ taken from Parker et al. (2005). North is up and east is to the left. Stars observed with SALT are marked by circles, and stars with VIMOS spectra are marked by crosses. The central cluster Dolidze 25 is also marked. At $d = 4000$ pc, 0.15° corresponds to $\sim$ 10 pc.

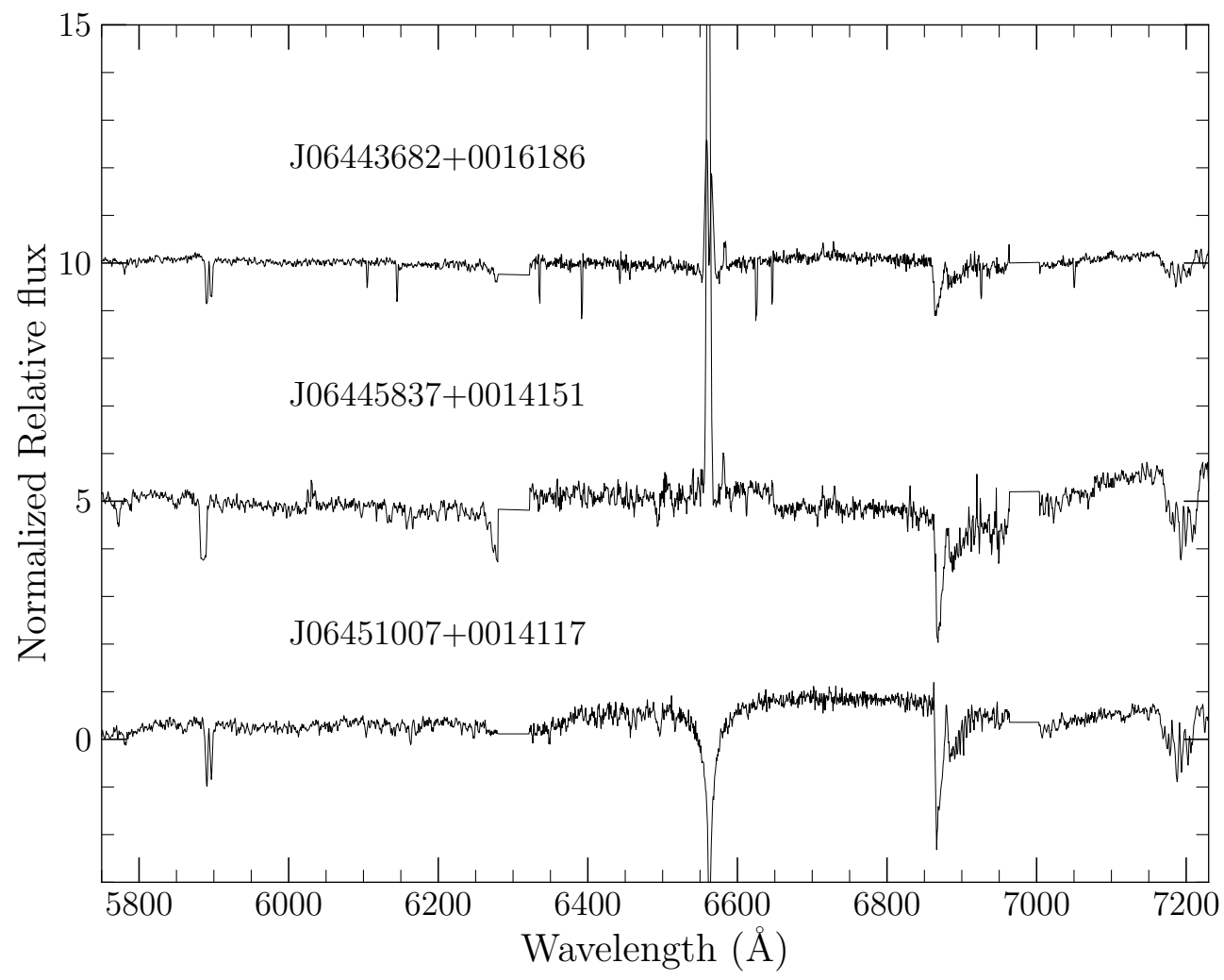

FIG. 2.— Normalized medium resolution longslit spectra of PMS candidates in Sh 2-284 observed using the SALT telescope. From top to bottom we show the spectra of an A9, F0 spectral type PMS stars, and a F6 star with no H$\alpha$ emission but infrared-excess. For the complete results see Table 2.

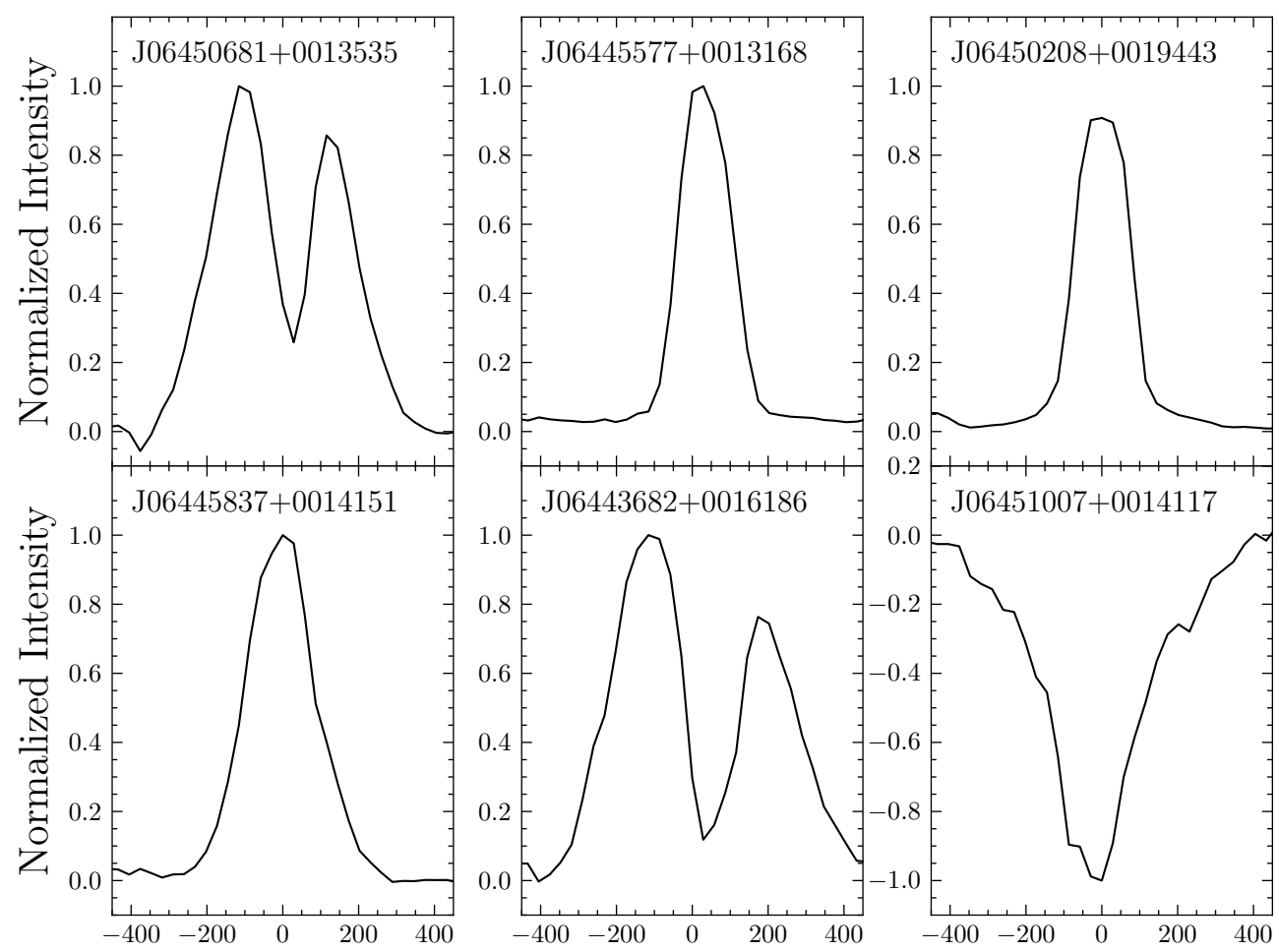

FIG. 3.— Normalized H$\alpha$ line profiles of the SALT spectra. The velocities of all stars with H$\alpha$ in emission is larger than 270 km s$^{-1}$, indicating clearly accretion in the line profile. The measured H$\alpha$ EW are given in Table 2.

### 2.1. *Basic analysis of the* SALT *data*

Equivalent widths (EW) of emission and absorption lines were measured using the IRAF task *splot*. Spectral types were derived from the EW of the (i) Mg II 5711 Å(ii) Mn II 6015 Å(iii) Ca I 5711 Å lines following the method devised by Hernández et al. (2004). Effective temperatures ($T_{\rm eff}$) were estimated using the spectral type-$T_{\rm eff}$ calibration of Kenyon & Hartmann (1995). H$\alpha$ line profiles are displayed in Fig. 3. Five stars show H$\alpha$ in emission. Combined with the fact that these stars all show near to mid-infrared excess resembling PMS stars, we take these five stars to be bonafide accreting PMS stars.

Model colors and bolometric corrections were calculated using Castelli & Kurucz (2004) model spectra at the appropriate $T_{\rm eff}$ and $Z$. We assumed the logarithm of surface gravity $\log g = 4.0 \pm 1.0$. Where available, $BV$ photometry from Delgado et al. (2010) was used to determine the reddening $E(B-V)$ and the logarithm of luminosity ($\log L$). We used a reddening law with $R_V = 3.1$ to determine the absolute visual extinction $A_V$. We adopted a distance $d = 4000 \pm 400$ pc derived by Cusano et al. (2011) and Delgado et al. (2010) for calculating the luminosity. Two stars (J06445578+001368 and J06450208+0019443) that do not have Delgado et al. (2010) optical photometry, archival $B-V$ colors (Monet et al. 2003), and 2MASS $J$-band magnitudes were used to derive the extinction and luminosity respectively. We assumed that the effect of any infrared excess in the $J$-band is comparatively small. Table 2 summarizes the fundamental parameters determined for these stars using SALT spectra.

### 2.2. VLT/VIMOS *data*

Additional data of PMS stars in Sh 2-284 was provided in Cusano et al. (2011). These authors obtained multislit spectroscopy of $\sim$ 900 objects covering a $30 \times 30$ arcmin$^2$ area (see Fig. 1) using the Visual & Multi-Object Spectrograph (VIMOS) on the Very Large Telescope (VLT). They identified 23 bonafide PMS stars from this sample based on H$\alpha$ emission, spectral type, and near-infrared excess criteria. 17 show infrared-excess resembling Class II sources, whilst the remaining objects resemble Class III sources (i.e. weakly accreting PMS stars). The location of these stars is shown in Fig. 1. Three stars (J06443682+0016186, J06450208+0019443, J06450681+0013535) also have SALT spectra.

The spectral types and fundamental parameters for 22 stars were determined by Cusano et al. (2011) using the method devised by Hernández et al. (2004). One star (J06452476+0013360) shows no absorption features which means that it did not allow for spectral classification. We adopt these results which are listed in Table 2. We compared the spectral parameters for the three stars in common with our SALT observations, and we found that spectral types agreed within 2 sub-classes, which is the average error. The agreement is expected as both spectral type determinations use similar line indicators and distances adopted by the authors are the same as the values used in the analysis of the SALT spectra. The logarithm of luminosity of one overlapping star (J06443682+0016186) measured using near-infrared $J$-band magnitude is higher by 0.25 dex to the value measured using Cusano et al. (2011) photometry. This is attributed to the near-infrared excess. For this star, we adopt the logarithm of luminosity determined by Cusano et al. For all other parameters determined using both SALT and VIMOS spectra, we adopt the values estimated from the SALT spectra.

The overall sample consists of 24 bonafide PMS stars in Sh 2-284 with spectral types ranging from mid A- early G for which there are observed parameters, which are necessary to infer accretion properties.

## 3. RESULTS

### 3.1. *Masses and Ages*

Using the measured effective temperatures and luminosities, we place the PMS stars in an Hertzsprung-Russell (H-R) diagram Fig. 4. Overlaid are the non-accreting single star isochrones and tracks calculated by Bressan et al. (2012) at $Z = 0.004$. The masses and ages are measured by interpolating their positions with respect to tracks and isochrones, respectively. The derived masses and ages are listed in Table 3. The differences that arise due to variations between different sets of tracks and isochrones are discussed in Appendix A.

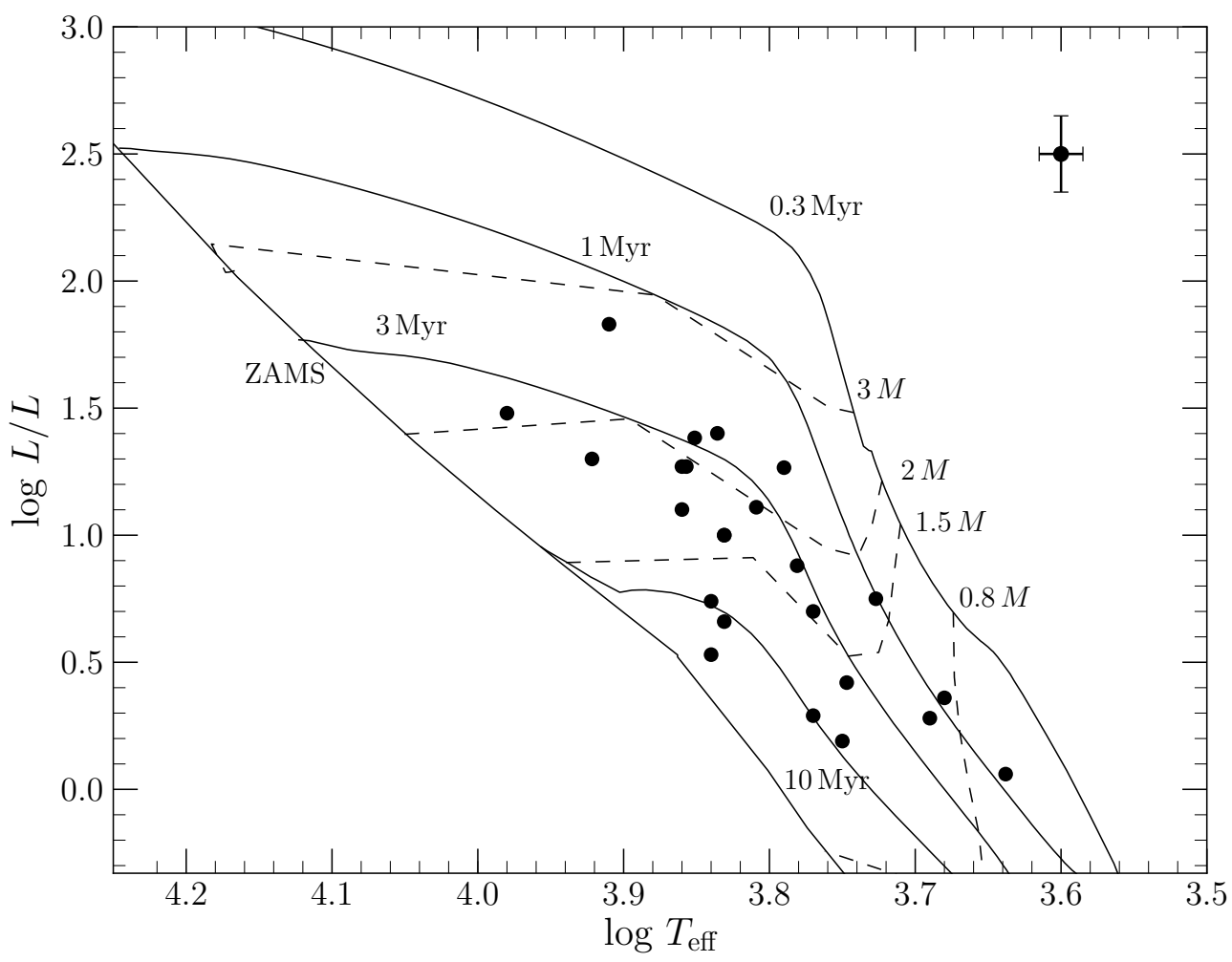

FIG. 4.— Hertzsprung-Russell diagram of Sh 2-284 PMS stars. The solid vertical lines from right to left are $Z = 0.004$ PMS isochrones (Bressan et al. 2012) at 0.3, 1, 3 and 10 Myr, and the zero-age main sequence respectively. The dashed horizontal lines from bottom up are the 0.8, 1.5, 2, and 3 $M_\odot$ tracks. The average error bars in $\log T_{\rm eff}$ and $\log L$ are shown in the top-right hand corner. The $\log T_{\rm eff}$ and $\log L$ of the majority of our PMS stars have been taken from Cusano et al. (2011). See Table 3 for the interpolated ages, and masses.

The mass range (0.9 - 2.6 $M_\odot$) places a majority of the stars in the intermediate mass T Tauri or Herbig Ae type range. These stars are very similar to accreting Classical T Tauri stars and their $\dot{M}_{\rm acc}$ can be reliably measured using their H$\alpha$ emission line luminosity (Calvet et al. 2004). The median error on the masses is $\lesssim 0.1\,M_\odot$. The ages determined from isochrones span a range of 1 - 15 Myr. The PMS stars in the central cluster Dolidze 25 have ages around 2 - 3 Myr, including the newly identified PMS stars using the SALT spectra (see Fig. 1). Those located at the edges of the surrounding nebulous bubbles (see Puga et al. 2009 for a detailed description of the nomenclature) have younger ages, whereas those at the rim have older ages. These results are in agreement with the near-mid infrared study of Puga et al. (2009) and spectroscopic-photometric work of Cusano et al. (2011). Puga et al. (2009) interpret this as evidence that star formation was triggered by a previous generation of ionizing stars in these regions, which has lead to the spread of PMS ages inferred from the H-R diagram. Their argument was supported by a first-principle analysis of the gas dynamics within the region.

PMS membership of these stars is not suspect, as all objects display infrared excesses resembling PMS stars, and there is little evidence for mid-A to early-G main sequence stars having H$\alpha$ EW $< -5$ Å. The older stars are located in the nebulous B2 and B3 regions (see Fig. 1). Therefore, the possibility that they are background stars detected through nebulosity is small, as they would have been significantly more extincted than the other PMS stars. Moreover, one early G star (J06451318+0018307), with an estimated age $> 10$ Myr, also displays Li in absorption in high signal-to-noise high resolution VIMOS spectra. Li $\lambda\,6708$ Å absorption is a reliable youth indicator in stars later than G (e.g. Duncan & Jones 1983; Favata et al. 1998). We defer from using Li EW to estimate age, as the effect of lower metallicity in the spectral range under study on the lithium boundary is unclear.

TABLE 2
Physical parameters of PMS stars in Dolidze 25

| ID (2MASS) | Spectral type | EW (Hα) (Å) | $A_V$ (mag) | $\log T_{\rm eff}{}^a$ (K) | $\log L^b$ ($L_\odot$) |
|---|---|---|---|---|---|
| SALT spectra | | | | | |
| J06443682+0016186 | A9 | $-8.98 \pm 0.4$ | $1.62 \pm 0.25$ | 3.87 | $1.10^c$ |
| J06445577+0013168 | F2 | $-8.23 \pm 0.3$ | $1.93 \pm 0.15$ | 3.84 | 1.40 |
| J06445837+0014151 | F0 | $-26.58 \pm 1.3$ | $2.45 \pm 0.15$ | 3.85 | 1.38 |
| J06450208+0019443 | A6 | $-43.56 \pm 2.1$ | $2.23 \pm 0.30$ | 3.92 | $1.53^c$ |
| J06450681+0013535 | F9 | $-1.46 \pm 0.5$ | $2.06 \pm 0.15$ | 3.78 | 1.26 |
| VIMOS spectra[d] | | | | | |
| J06443682+0016186[e] | A8V | $-8.4 \pm 0.2$ | 1.9 | 3.88 | 0.98 |
| J06450208+0019443[e] | A5V | $-8.05 \pm 0.1$ | 2.44 | 3.91 | 1.27 |
| J06450681+0013535[e] | G0V | $-1.9 \pm 0.1$ | 2.0 | 3.77 | 1.25 |
| J06451701+0022077 | A0V | $-10.3 \pm 0.4$ | 3.82 | 3.98 | 1.48 |
| J06450971+0014121 | A5V | $-14.0 \pm 0.2$ | 2.11 | 3.91 | 1.83 |
| J06443291+0023546 | F0V | $-23.8 \pm 0.4$ | 3.1 | 3.86 | 1.27 |
| J06443788+0021509 | F0V | $-2.8 \pm 0.1$ | 2.28 | 3.86 | 1.27 |
| J06443541+0019093 | F2V | $-4.2 \pm 0.2$ | 3.0 | 3.83 | 1.0 |
| J06443827+0019229 | F2V | $-11.7 \pm 0.25$ | 1.3 | 3.83 | 0.66 |
| J06451727+0023454 | F2V | $-16.58 \pm 0.3$ | 2.69 | 3.83 | 1.0 |
| J06450274+0018077 | F3V | $-33.7 \pm 0.25$ | 3.14 | 3.84 | 0.74 |
| J06451616+0022238 | F3V | $-5.3 \pm 0.25$ | 3.01 | 3.84 | 0.53 |
| J06443294+0010528 | F4V | $-139.28 \pm 1.23$ | 3.64 | 3.81 | 1.11 |
| J06444496+0019335 | F8V | $-8.8 \pm 0.15$ | 2.52 | 3.78 | 0.88 |
| J06444602+0019182[f] | G0V | $-73.2 \pm 0.9$ | 5.63 | 3.77 | 0.7 |
| J06451318+0018307 | G0V | $-40.5 \pm 0.5$ | 2.64 | 3.77 | 0.29 |
| J06451841+0022189 | G4V | $-5.2 \pm 0.3$ | 1.05 | 3.75 | 0.19 |
| J06444936+0020245 | G5V | $-14.87 \pm 0.3$ | 1.7 | 3.75 | 0.42 |
| J06452454+0022448 | G8V | $-21.18 \pm 0.2$ | 2.33 | 3.73 | 0.75 |
| J06450075+0013356 | K0V | $-31.69 \pm 0.5$ | 1.89 | 3.69 | 0.28 |
| J06444714+0013320[f] | K2V | $-20.5 \pm 0.5$ | 2.15 | 3.68 | 0.36 |
| J06452774+0015514 | K4V | $-29.0 \pm 1.3$ | 2.77 | 3.64 | 0.06 |

[a] Average error in $\log T_{\rm eff}{}^a$ is around $\lesssim 0.05$ dex due to uncertainty in spectral class. The spectral class classification is uncertain within 1.5 subclasses from SALT spectra, and 1 subclass from VIMOS spectra.
[b] Average error in $\log L$ is $\sim 0.15$ dex mainly due to uncertainty in $d \sim 500$ pc.
[c] $\log L$ was calculated using 2MASS $J$-band magnitude and archival (Monet et al. 2003) $B-V$ colours.
[d] Taken from Table 4 in Cusano et al. (2010), where the quoted average errors in spectral type is one subclass, $A_V \sim 0.1$ mag.
[e] Star also has SALT longslit spectra.
[f] Star not detected in 2MASS, coordinates from Cusano et al. (2010).

Errors in spectral type determination larger than two-three sub-classes for these stars could lead to incorrect temperature determinations, but also extinction and bolometric corrections, and thereby luminosity. However, the two effects roughly cancel leading to small absolute age errors (see Hartmann 2001). Moreover, any binary companions would roughly double the measured luminosity, while keeping the temperature the same, leading to younger ages. Another source of error may involve the stellar isochrones themselves. Differences in input physics imply that absolute ages derived using different isochrones could differ by as much as a factor of 2. We tested this by comparing the ages derived using the Bressan et al. (2012) isochrones from the PISA database (Tognelli et al. 2011) used for age determination by Cusano et al. (2011) in Appendix A. We find that both isochrones gives similar absolute age values. Overall, we take $\lesssim 1$ Myr as an average error for the interpolated age.

We noticed that a significant fraction of stars with estimated ages higher than average for the region have masses of $\sim 1\,M_\odot$. This leads us to suggest that the age spread could possibly be due to errors in birthline calculations for G-type stars (Hartmann 2003), which may lead to a systematic overestimation of absolute stellar ages. Therefore, any ages derived can only be accurate in a relative sense. Any such birthline effect is circumvented by interpolating the ages of solar metallicity PMS stars used for comparison using the same isochrones and following the same interpolation technique. This is described further in Appendix A.

### 3.2. *Accretion luminosity*

In the current PMS evolutionary paradigm, stars accrete mass from a surrounding circumstellar disk. The disk is thought to be curtailed at some inner radius ($R_{\rm in}$) by the stellar magnetosphere, and gas in the disk is accreted via the magnetic field lines. The gravitational en-ergy is released, as material falls onto the stellar surface creating an accretion shock, leading to excess continuum emission, which is particularly evident in the ultraviolet wavelength range. The ionization and recombination of gas during this process leads to strong Balmer line emission (notably in H$\alpha$). In this scenario, the bolometric accretion luminosity ($L_{\rm acc}$) can be estimated from the re-radiated energy, which can be measured from the H$\alpha$ line luminosity ($L_{\rm H\alpha}$).

We measure $L_{\rm H\alpha}$ by adapting the method of Mohanty et al. (2005) (see also Costigan et al. (2014) for an application to higher-mass stars). The flux from Castelli & Kurucz (2004) model spectra (Z = 0.004) for the nearest point in ($T_{\rm eff}$,log $g$) plane in 6554 - 6572 Å range is taken as continuum flux. We adopt for all stars log $g$ = 4.0.

The continuum flux is multiplied by the H$\alpha$ EW and $4\pi R^2$ to obtain $L_{\rm H\alpha}$. Veiling is estimated by comparing with Pickles (1998) non-accreting stellar spectra. In the range with measurable absorption lines ($>5800$ Å), veiling is negligible. This is in good agreement with the results of Calvet et al. (2004) for stars in a similar wavelength and mass range, but using high-resolution spectra over a much larger wavelength range. We adopt the empirical $L_{\rm H\alpha}$-$L_{\rm acc}$ relation used by De Marchi et al. (2010);

$$\log\, L_{\rm acc} = \log\, L_{\rm H\alpha} + 1.72 \pm 0.47 \qquad (1)$$

The relation was derived by comparing $L_{\rm acc}$ estimated from the UV continuum-excess with $L_{\rm H\alpha}$ of 0.5 - 2 $M_\odot$ mass stars (Dahm 2008). log $L_{\rm acc}$ estimated this way includes an uncertainty of factor 0.5, which is the dominant error in the measured $\dot{M}_{\rm acc}$. The $L_{\rm H\alpha}$-$L_{\rm acc}$ used by De Marchi et al. (2010) agrees within errors with that in a similar range derived by Mendigutía et al. (2011). We stress that the $L_{\rm H\alpha}$-$L_{\rm acc}$ relation was adopted with a view to reduce any sources of systematic errors between our results to those in the LMC. A comparison of the method used to calculate $L_{\rm acc}$ with respect to traditional H$\alpha$ line flux measurements is presented in Appendix B.

### 3.3. *Mass accretion rate*

$\dot{M}_{\rm acc}$ can be estimated from the $L_{\rm acc}$ using the free-fall equation;

$$\dot{M}_{\rm acc} = \frac{L_{\rm acc} R_*}{{\rm GM}_*}\left(\frac{R_{\rm in}}{R_{\rm in} - R_*}\right). \qquad (2)$$

Here $M_*$, $R_*$ are the stellar mass and radius respectively. The inner radius ($R_{\rm in}$) is uncertain and depends on the coupling of the accretion disk with the magnetic field lines. We follow Gullbring et al. (1998) and take $R_{\rm in} = 5{\pm}2\, R_\odot$ (Vink et al. 2005). Using the stellar mass and radius measured from the H-R diagram, we estimate the $\dot{M}_{\rm acc}$ for each star. The accretion properties are tabulated in Table 3.

As a sanity check on the estimated $\dot{M}_{\rm acc}$, we compared whenever possible with archival $U$-band photometry of Delgado et al. (2010). $\dot{M}_{\rm acc}$, or more directly the $L_{\rm acc}$ can be measured reliably from $U$-band photometry. Comparison of the observed $U$-band magnitude with a template magnitude comprises a direct measure of the excess $U$-luminosity caused by the emission shock which can be translated into $L_{\rm acc}$ (Gullbring et al. 1998).

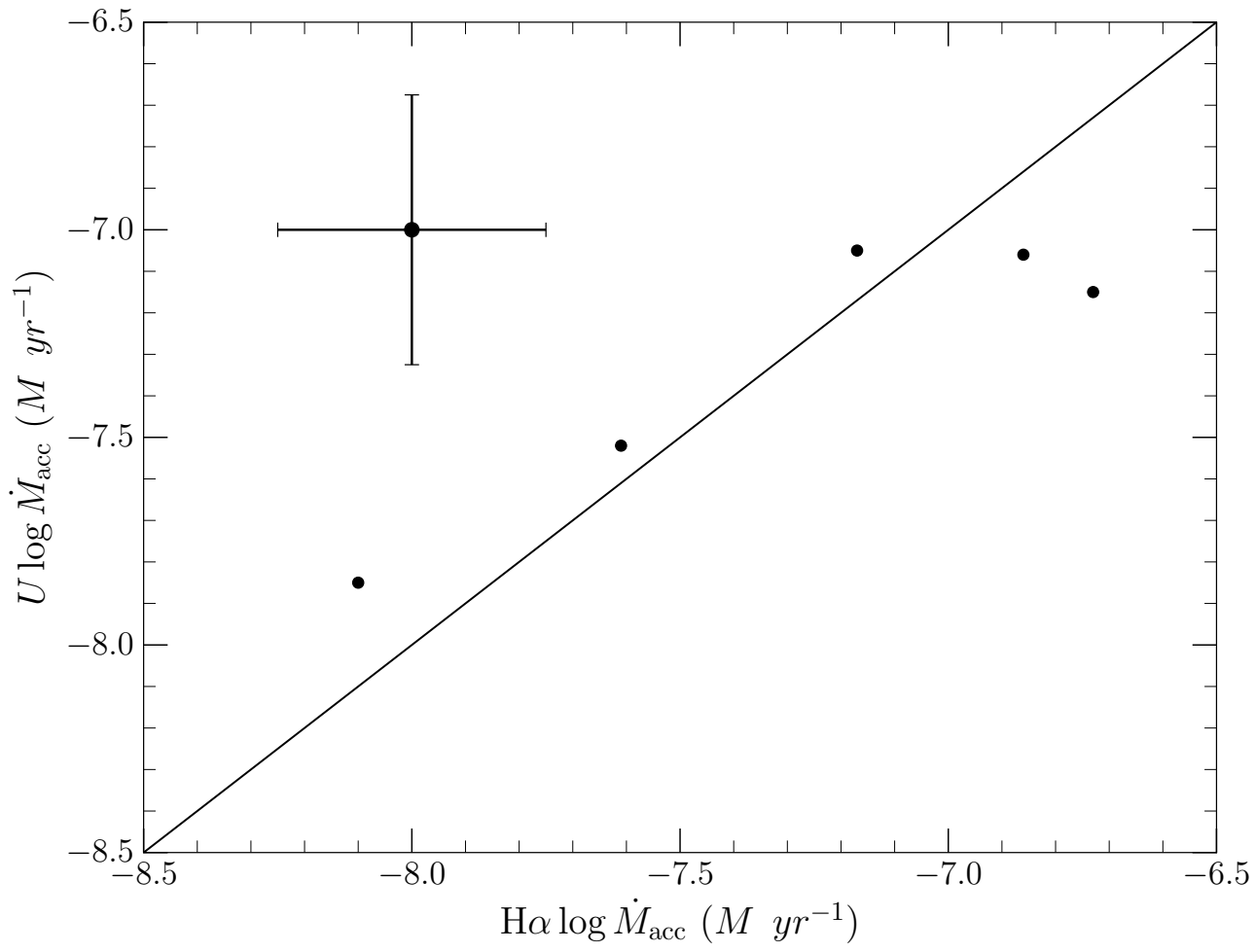

Fig. 5.— Comparison of $\dot{M}_{\rm acc}$ measured from H$\alpha$ spectroscopy (abscissa) with archival $U$-band photometry (ordinate). The average errors are shown in the top-left hand corner. Although the $U$-band sample is small, the comparison suggests consistency between mass-accretion rates determined from H$\alpha$ and the $U$-band, i.e. H$\alpha$ is most likely a good indicator for mass-accretion phenomena.

Five stars out of the total 24 have archive $U$-band photometry in Delgado et al. (2010). The excess $U$-band luminosity ($L_{U,\rm excess}$) of these stars, thought to be due to accretion is estimated according to the formula

$$L_{U,\rm excess} = 4\pi{\rm W}(d^2 f_{\rm obs} - R_*^2 f_{\rm mod}) \qquad (3)$$

(Romaniello et al. 2002). $f_{\rm mod}$ and $f_{\rm obs}$ are the model and observed fluxes in the $U$-band respectively. W is the width of the $U$-band filter. The $L_{U,\rm excess}$ is translated into $L_{\rm acc}$ following the relation of Gullbring et al. (1998)

$$\log\, L_{\rm acc} = 1.09\, \log\, L_{U,\rm excess} + 0.98. \qquad (4)$$

Using eq. (2), we compute $\dot{M}_{\rm acc}$. A comparison of the $\dot{M}_{\rm acc}$ derived using $U$-band photometry and H$\alpha$ line luminosity are presented in Fig. 5. We find that the difference is not greater than the mean error (0.5 dex), suggesting that the $\dot{M}_{\rm acc}$ is reasonably well constrained for our purposes.

### 3.4. *Disk properties*

We have shown that all 24 PMS stars in Sh 2-284 have H$\alpha$ in emission, and for a subset the accretion rates derived from H$\alpha$ agree with those derived from the U-band (see also Kalari et al. 2014, in prep.). This demonstrates to a reasonable degree that these stars are accreting. Most of these stars also have near-infrared 2MASS *JHKs* magnitudes (Cutri et al. 2003), and *Spitzer* imaging up to 8$\mu$m (Puga et al. 2009). In this section, we analyze the disk properties of the PMS objects using infrared photometry.

The slope of the SED in the infrared,

$$\alpha_{\rm IR} = \frac{d\log(\lambda F_\lambda)}{d\log\lambda}, \qquad (5)$$

at $\lambda > 3\,\mu$m is used to diagnose the evolutionary stage of the disk-star system (Lada 1987). We adopt the classification scheme of Greene et al. (1994) to distinguish between systems with protostellar disks (Class I), optically thick disks (Class II), or no circumstellar excess (Class III). $\alpha_{\rm IR}$ was derived by fitting the *Spitzer* magnitudes at 3.6, 4.5, 5.8, and 8$\mu$m. The resultant $\alpha_{\rm IR}$ values are given in Table 3. The disk SEDs are shown in Fig. 6. Five stars do not have photometry in any band. For these stars, we fitted the available *Spitzer* magnitudes to the disk models of Robitalle et al. (2006) to classify the disk evolutionary state.

TABLE 3
Accretion and disc properties of PMS stars in Dolidze 25

| No. | ID (2MASS) | Mass ($M_\odot$) | Age (Myr) | $\log L_{\rm acc}$ ($L_\odot yr^{-1}$) | $\log \dot{M}_{\rm acc}$ ($M_\odot yr^{-1}$) | $\alpha_{\rm IR}$ | Class |
|---|---|---|---|---|---|---|---|
| 1 | J06443291+0023546 | 1.79 | 4.0 | 0.28 | −6.92 | −0.27 | II |
| 2 | J06443294+0010528 | 1.83 | 3.5 | 0.95 | −6.24 | −0.28 | II |
| 3 | J06443541+0019093 | 1.57 | 5.6 | −0.71 | −7.93 | −2.02 | III |
| 4 | J06443682+0016186 | 1.6 | 5.5 | −0.26 | −7.5 | −0.89 | II |
| 5 | J06443788+0021509 | 1.79 | 3.9 | −0.65 | −7.85 | - | III |
| 6 | J06443827+0019229 | 1.26 | 10.9 | −0.6 | −7.9 | −1.41 | II |
| 7 | J06444496+0019335 | 1.75 | 3.3 | −0.46 | −7.69 | −0.97 | II |
| 8 | J06444602+0019182 | 1.57 | 3.8 | 0.29 | −6.96 | - | III |
| 9 | J06444714+0013320 | 0.91 | 0.9 | −0.6 | −7.61 | −0.76 | II |
| 10 | J06444936+0020245 | 1.36 | 4.3 | −0.67 | −7.95 | 0.59 | I |
| 11 | J06445577+0013168 | 2.08 | 2.7 | −0.04 | −7.2 | 0.38 | I |
| 12 | J06445837+0014151 | 2.0 | 3.0 | 0.38 | −6.79 | 0.70 | I |
| 13 | J06450075+0013356 | 1.05 | 1.5 | −0.49 | −7.61 | −1.25 | II |
| 14 | J06450208+0019443 | 1.75 | 4.1 | 0.48 | −6.82 | 0.92 | I |
| 15 | J06450274+0018077 | 1.3 | 9.5 | −0.07 | −7.36 | −0.23 | II |
| 16 | J06450681+0013535 | 2.16 | 2.0 | −0.92 | −8.07 | −0.3 | II |
| 17 | J06450971+0014121 | 2.62 | 1.5 | 0.53 | −6.66 | - | II |
| 18 | J06451318+0018307 | 1.15 | 9.3 | −0.38 | −7.7 | −1.09 | II |
| 19 | J06451616+0022238 | 1.23 | 14.8 | −1.09 | −8.46 | - | II |
| 20 | J06451701+0022077 | 2.0 | 7.5 | −0.07 | −7.45 | - | II |
| 21 | J06451727+0023454 | 1.57 | 5.6 | −0.11 | −7.33 | −0.32 | II |
| 22 | J06451841+0022189 | 1.12 | 8.5 | −1.36 | −8.68 | −0.52 | II |
| 23 | J06452454+0022448 | 1.66 | 1.0 | −0.18 | −7.35 | −0.98 | II |
| 24 | J06452774+0015514 | 0.56 | 0.9 | −0.8 | −7.66 | −0.77 | II |

Out of the 19 accreting PMS stars that have *Spitzer* photometry at all observed wavelengths, we find that 18 have SED slopes resembling Class I/II objects, with one star having $\alpha_{\rm IR}$ resembling a Class III object. Three stars do not have 8 $\mu$m photometry (J06443788+0021509, J06444602+0019182), of which two have best-fit disk models suggestive of either flat or sources devoid of any infrared excess, and one (J06451701+0022077) suggestive of a dusty disk. Two stars that do not have photometry at either 4.5, or 5.8 $\mu$m (J06451701+0022238, J06450971+0014121), have best-fit models suggestive of Class II objects. Overall, most of the PMS star sample have infrared-excess akin to circumstellar disks, verifying that we deal with genuine accreting PMS stars.

## 4. DISCUSSION

We present spectroscopic estimates of accretion rates of 24 low-$Z$ PMS stars in a stellar mass range between 1 - 2 $M_\odot$, having a median age of 3.5 Myr, and a median accretion rate of $10^{-7.5}\,M_\odot yr^{-1}$. The majority of our sample (21 out of 24 stars) shows evidence for circumstellar disks, and all stars with literature $U$-band photometry (5 out of 24) display an UV excess, confirming the PMS nature of the sample. Based on our results, we discuss in the following section the relation between $L_{\rm acc}$ and stellar luminosity; and the observed dependence of the estimated $\dot{M}_{\rm acc}$ on stellar mass and age. We also compare our results with literature estimates of accretion properties of $Z_\odot$ Galactic and $1/3Z_\odot$ LMC PMS stars to examine if $\dot{M}_{\rm acc}$ estimates are a function of $Z$ within the observed $Z$ range.

### 4.1. $L_{\rm acc}$ *vs.* $L_*$

The detection limit on $\dot{M}_{\rm acc}$ are dependent on the detection limits of the measured $L_{\rm acc}$. This is demonstrated in the $L_{\rm acc}$ vs. $L_*$ distribution in Fig. 7. We find that $L_{\rm acc} < L_*$, i.e., we did not identify any *continuum* stars (Hernández et al. 2004) in our sample, leading to an upper boundary in the $\dot{M}_{\rm acc}$-$M_*$ plane, possibly leading to a scarcity of high $\dot{M}_{\rm acc}$. However, it is thought that continuum stars form only a very small fraction of the total population in well-studied regions such as Taurus (Hartmann 2008).

The lower detection limits have been calculated from $L_*$ and $M_*$ of stars lying on the 3 Myr isochrone with $T_{\rm eff} = 5000$, 6000, 7000, 8000, 10000 and 12000 K. We assume a H$\alpha$ EW = −2 Å to calculate their accretion luminosity. The limit is near the point where accretion is not detectable in H$\alpha$ line emission. We note that at $L_{\rm acc}$ lower than this boundary it would be impossible to detect accretion using conventional means such as line emission or UV-excess, since H$\alpha$ is the most sensitive indicator to tiny accretion rates. The lower boundary of $L_{\rm acc}$ is within the error bars of detections at $L_* < 1\,L_\odot$, at approximately 1.5 $M_\odot$. The PMS stars identified by Cusano et al. (2011) which form the bulk of our sample stars, were identified by the authors of that paper as those stars showing H$\alpha$ emission out of 1506 spectra observed in the region. Stars were rejected as nonmembers based on their luminosity classification, and infrared colors. It is plausible to expect that the Cusano et al.(2011) PMS sample would contain stars having lower $L_{\rm acc}$ at higher masses if they existed. This suggests that our sample contains PMS stars at the lowest measurable $\dot{M}_{\rm acc}$ within this range. One star lies at the boundary of the lower detection limit (J06450681+0013535) and has a measured H$\alpha$ EW of $-1.5$ Å, lower than the assumed threshold.

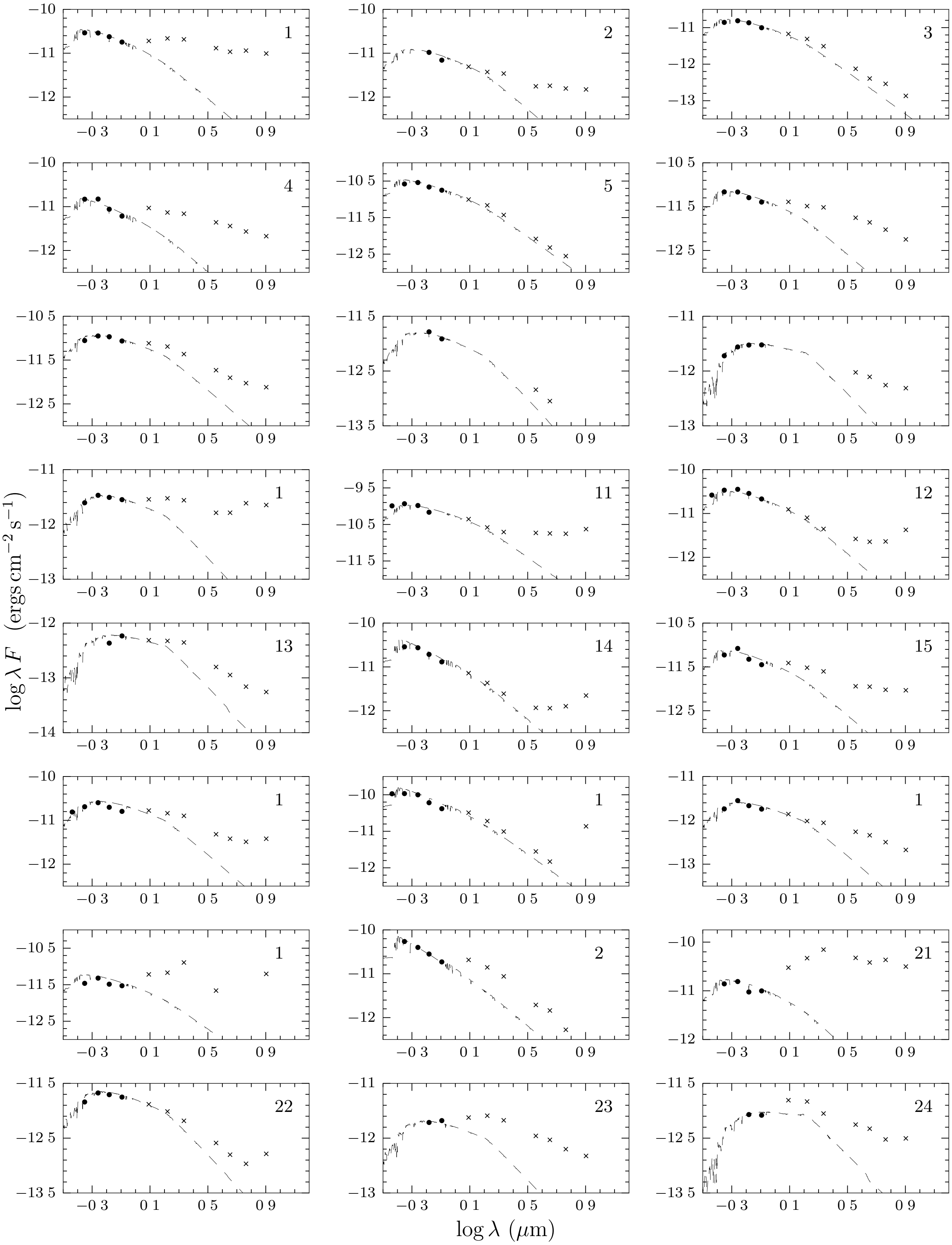

FIG. 6.— The spectral energy distribution of Sh 2-284 pre-main sequence stars. The number in the top-right corner corresponds to the index number in Table 3. Dashed lines are the Kurucz & Castelli (2004) model spectra corresponding to the determined spectral parameters in Table 2. Dots are the un-reddened *U,B,V,R,I* fluxes from Delgado et al. (2010) or Cusano et al. (2011). Crosses are *JHK*s fluxes from 2MASS (Cutri et al. 2003) and *Spitzer* IRAC fluxes at 3.6, 4.5, 5.8, and 8.0 $\mu$m.

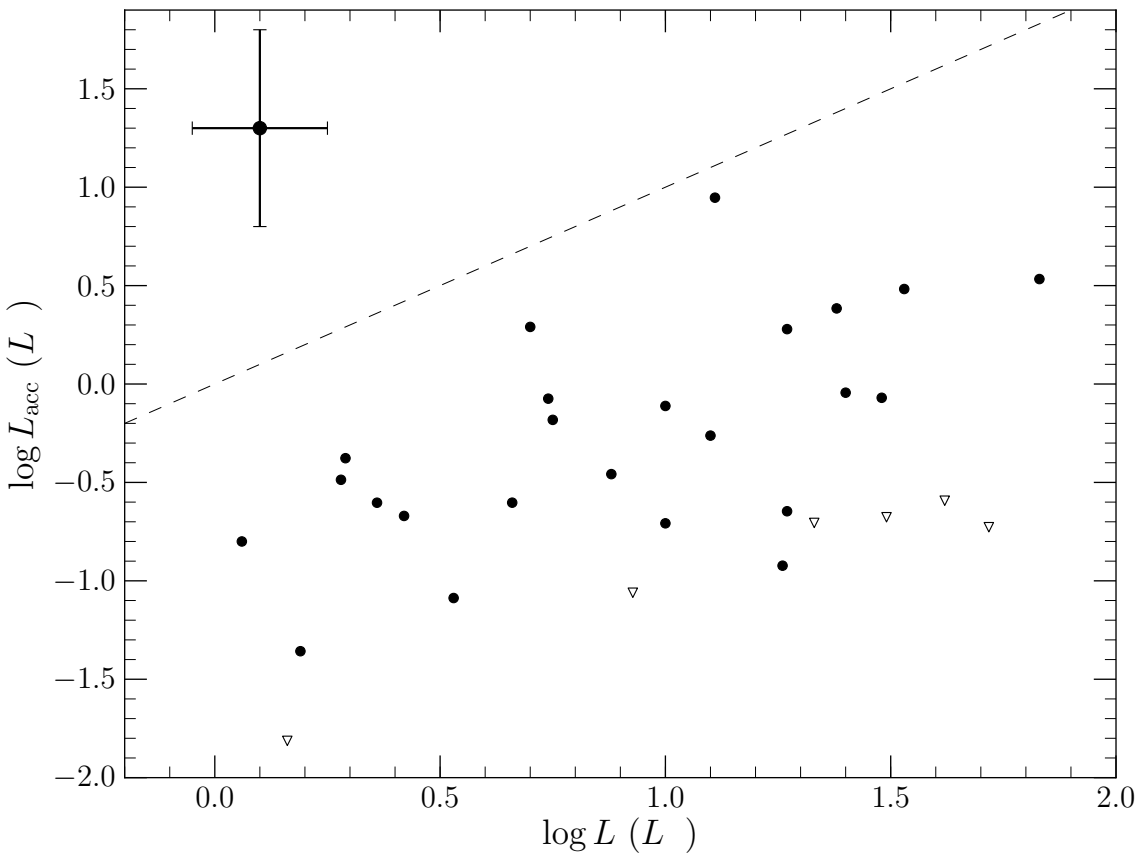

FIG. 7.— $L_{\rm acc}$ as function of $L_*$. The dashed line displays the $L_* = L_{\rm acc}$ relation. The inverted triangles give the detection limits calculated for six different $T_{\rm eff}$ for model $L_*$ at 3 Myr. The upper left corner displays the mean error bars. The $L_{\rm acc}$ are given in Table 3.

### 4.2. $\dot{M}_{\rm acc}$ *as a function of stellar mass and age*

$\dot{M}_{\rm acc}$ are plotted as a function of $M_*$ in Fig. 8. We find the best-fit power law relationship across the 1 - 2 $M_\odot$ range to be $\dot{M}_{\rm acc} \propto {M_*}^{2.4\pm0.35}$. There is a spread of $\sim$ 2 dex in $\dot{M}_{\rm acc}$ at any given $M_*$. The spread can partially be related to the intrinsic distribution of ages at any given mass, as $\dot{M}_{\rm acc}$ vary with age. But the scatter is also found in coeval populations, suggesting that there are other factors beyond mass and age controlling accretion rates (Hartmann et al. 2006).

We employed survival analysis linear regression using ASURV (Lavalley et al. 1992) to include lower and higher limits in accretion rates. We find that including lower limits leads to a steeper slope, but within the error bar. The inclusion of upper limits leads to a shallower slope, but within 0.1 dex of the slope calculated. This suggests that the absence of low-mass high-accretion rates, or high-mass low-accretion rates is not statistically significant in our regression fitting, and that the given distribution statistically represents a general trend of increasing $\dot{M}_{\rm acc}$ with $M_*$. The index $\alpha$ measured is close to the Galactic median $\sim 2$, but deviant from the value of $\sim 1$ claimed in metal-poor PMS stars in the LMC (Spezzi et al. 2012).

The ages of stars in the Sh 2-284 region are around 3 Myr in the central cluster, but vary in the surrounding nebula (Puga et al. 2009). The age distribution of Sh 2-284 is reflected in the $\dot{M}_{\rm acc}$-Age plane (Fig. 8). We fit the orthogonal distance regression to account for errors on both $\dot{M}_{\rm acc}$ and $t_*$. We find the power-law index $\eta$ to be $-0.7\pm0.4$, similar to the slope found in studies of $Z_\odot$ PMS stars, and as expected in line with viscous disk evolution at these masses (Sicilia-Aguila et al. 2006; Manara et al. 2012).

Overall, a comparison of Sh 2-284 PMS stars with the literature accretion rates of $1/3\,Z_\odot$ LMC and $Z_\odot$ Galactic PMS stars is shown in Fig. 8. The Sh 2-284 PMS stars occupy similar ranges to $Z_\odot$ PMS stars in both the $\dot{M}_{\rm acc}$-$M_*$ and $\dot{M}_{\rm acc}$-$t_*$ plane. We find an absence of LMC PMS stars at low $\dot{M}_{\rm acc}$ which suggest that the overall slopes measured by Spezzi et al. (2012) may be significantly affected by detection limitations. This could be due to the difficulty of observing low-mass stars in the distant LMC. We also find an absence of stars with $\dot{M}_{\rm acc} > 10^{-7}\,M_\odot yr^{-1}$ at ages beyond 5 Myr in our sample, a region of the plane that is populated by points from the LMC.

### 4.3. *Is* $\dot{M}_{\rm acc}$ *a function of* $Z$*?*

We have shown that the $\dot{M}_{\rm acc}$ in Sh 2-284 and Galactic solar-$Z$ star forming regions follow similar distributions in the $\dot{M}_{\rm acc}$-$M_*$ and $\dot{M}_{\rm acc}$-$t_*$ plane. The PMS star sample in the LMC of Spezzi et al. (2012) have shallower slopes, which may be affected by detection limits. However, as discussed by Spezzi et al. (2012), the median $\dot{M}_{\rm acc}$ at any given age in a defined mass range (1-2 $M_\odot$) is a robust tool for comparing any differences in $\dot{M}_{\rm acc}$ between star-forming regions. To investigate this we tabulate the median $\dot{M}_{\rm acc}$ of 1 - 2 $M_\odot$ PMS stars in known star-forming regions in Table 4. The median age for the given sample is computed from the isochronal ages. To estimate the completeness of each sample, we calculate the expected fraction of stars within this mass range using the Kroupa (2001) IMF, and total cluster masses from Weidner & Kroupa (2006). The median age of each star-forming region is used to determine the expected fraction of accretion stars, $N_{\rm exp}$ (Fedele et al. 2010).

The median $\log \dot{M}_{\rm acc}$ of Sh 2-284 PMS stars is $-7.5$ dex. It is slightly larger than the Galactic mass-accretion rate by $\sim$ 0.3 dex in the IC 348, Trumpler 37 region and Cha II clouds, but similar to the values obtained in the Taurus-Auriga region and Lupus molecular clouds. In comparison, the median $\log \dot{M}_{\rm acc}$ reported by Spezzi et al. (2012) in an LMC PMS population having a median age of 6 Myr is $-7.1$ dex. Using Equation 7 from De Marchi et al. (2013), the median $\log \dot{M}_{\rm acc}$ expected in the SMC cluster NGC 602 at an age of 3 Myr for 1 - 2 $M_\odot$ is expected to be $\gtrsim -7.1$ dex.

Based on Table 4 we see that the median $\dot{M}_{\rm acc}$ in Sh 2-284 are comparable to Galactic values at similar ages. There is no systematic difference at an age of 3 Myr in the accretion rates of solar $Z$ or Sh 2-284 PMS stars based on these results. If we assume that our results are not significantly underestimated (see Section 4.2), it is challenging to explain physically how accretion is proceeding at much higher rates in much older stars in the LMC, especially considering that $Z_{\rm LMC} \sim 2\times Z_{\rm Sh\,2-284}$. It seems unlikely that $Z$ is the only quantity determining the accretion rates, and factors other than $Z$ may be responsible for the high $\dot{M}_{\rm acc}$ measured by the studies of De Marchi et al. (2010; 2011a;2011b;2013) and Spezzi et al. (2012).

There are practical considerations which might lead to higher measured $\dot{M}_{\rm acc}$ at low $Z$. The observable limit for the highest measurable $\dot{M}_{\rm acc}$ is generally given by $L_{\rm acc} < L_*$, owing to the difficulty in detecting *continuum* stars, and the lowest $\dot{M}_{\rm acc}$ at H$\alpha$ EW $< 1\,\mathring{A}$ beyond which

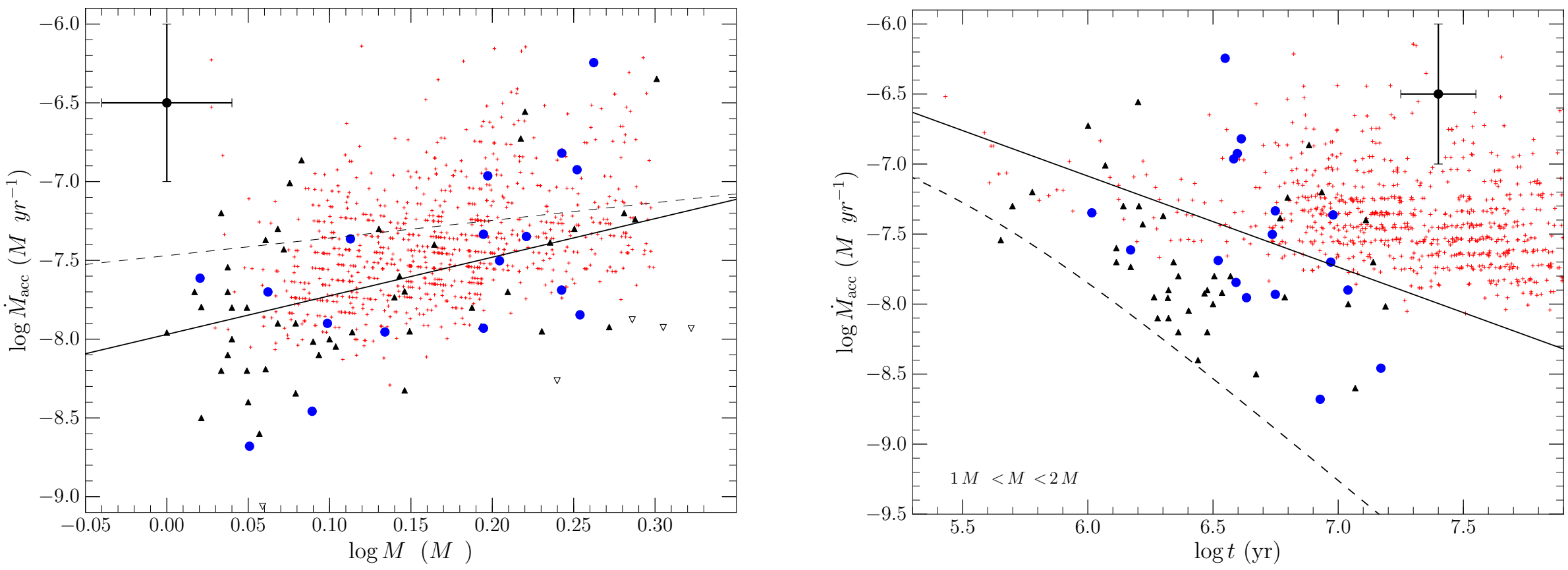

FIG. 8.— Distribution of $\dot{M}_{acc}$ as function of $M_*$ (left) and $t_*$ (right) shown as blue circles. The solid line is the best-fit regression slope in the 1 - 2 $M_\odot$ range, and has an index $\alpha = 2.4 \pm 0.35$ in the $\dot{M}_{acc}$-$M_*$ plane and index $\eta = 0.7 \pm 0.4$ in the $\dot{M}_{acc}$-$t_*$ plane. The dashed line (right) is the expected evolution of viscous disks following Hartmann et al. (1998), and left is the slope for LMC PMS stars reported in Spezzi et al. (2012). Black inverted triangles are the detection limits. Triangles are the data points for Galactic PMS taken from the studies of Natta et al. (2006); Herczeg & Hillenbrand (2008); Sicilia-Aguilar et al. (2010); Barentsen et al. (2011; 2013); Antoniucci et al. (2011); Manara et al. (2012); Alcále et al. (2014). Red crosses are taken from the studies of LMC PMS stars from Spezzi et al. (2012).

TABLE 4
MEDIAN $\dot{M}_{acc}$ OF PMS STARS IN THE 1 - 2 $M_\odot$ AT DIFFERENT $Z$

| Region | N | $N_{exp}$ | Age (Myr) | log $\dot{M}_{acc}$ ($M_\odot yr^{-1}$) | Reference |
|---|---|---|---|---|---|
| $Z = Z_\odot$ | | | | | |
| L1641 | 23 | - | 1 | −7.1 | Fang et al. (2009); Caratti o Garatti et al. (2012) |
| Orion nebula cluster | 41 | 75±10 | 1.5 | −7.65 | Manara et al.(2012) |
| $\rho$-Ophiuchus | 5 | 5±3 | 0.5-2[a] | −7.95 | Natta et al.(2006) |
| NGC 2264 | 7 | 15±4 | 3 | −7.8 | Barentsen et al.(2013) |
| Taurus-Auriga | 3 | 2±1 | 3 | −7.4 | Calvet et al.(2004) |
| IC 348 | 2 | 6±2 | 1.5-4[b] | −7.8 | Dahm (2008) |
| Lupus | 3 | - | 3 | −7.6 | Alcalá et al.(2014); Herczeg & Hillenbrand (2008) |
| Cha II | 7 | - | 3-4 | −7.9 | Antoniucci et al.(2011) |
| Trumpler 37 | 42 | (b) | 4 | −7.85 | Sicilia-Aguilar et al.(2010); Barentsen etal.(2011) |
| $Z = 0.4\,Z_\odot$ | | | | | |
| LMC Field3 Pop.2 | 146 | | 6 | −7.2 | Spezzi et al.(2012) |
| LMC Field3 Pop.1 | 96 | | 8 | −7.2 | Spezzi et al.(2012) |
| LMC Field2 | 325 | | 9 | −7.4 | Spezzi et al.(2012) |
| LMC Field1 | 490 | | 13 | −7.6 | Spezzi et al.(2012) |
| SN 1987 A | 104 | | 14 | −7.6 | De Marchi et al.(2010) |
| $Z = 0.2\,Z_\odot$ | | | | | |
| Sh 2-284 | 20 | | 3.5 | −7.5 | This work |
| $Z = 0.1\,Z_\odot$ | | | | | |
| NGC 346[d] | - | | 1 | −7.1 | De Marchi et al.(2011) |
| NGC 346[d] | - | | 20 | −7.7 | De Marchi et al.(2011) |

NOTE. — (a) $\rho$-Ophiuchus is oft-quoted with an age of 0.5 Myr (Natta et al. 2006), however the ages of the accreting PMS stars are thought to be as old as 2 Myr (Wiliking et al. 2008; Rigiliaco et al. 2010). (b) The total cluster mass of Trumpler 37 is debated, as studies of stellar content reveal a dearth of near solar mass members, and a deviant stellar mass function slope (Errman et al. 2012). (c) The age of IC 348 is debated, with initial studies suggesting an age as young as 1.5 Myr (Herbig 1998), but later revised to around 4 Myr (Mayne & Naylor 2007). (d) Taken from Figure 8 of Spezzi et al. (2012).

accretion is hard to measure. The lower limit leads to a threshold in $L_*$ (shown in Fig. 7). $L_*$ increases inversely in a non-monotonic fashion as a function of $Z$ according to stellar isochrones. This suggests that we are measuring stars of similar spectral type, but with different $L_*$ at different $Z$, with consequently higher $L_{\rm acc}$ and thereby $\dot{M}_{\rm acc}$ at lower $Z$. This also leads to raising the $L_{\rm acc}$ - $L_*$ upper limit shown in Fig. 7, and leading to a slightly higher $\dot{M}_{\rm acc}$. To test the difference that may be caused by this effect, we calculate the lower detection limit using solar $Z$ isochrones and use these to measure the median $\dot{M}_{\rm acc}$ for censored data using a Kapalan-Meier test. We find that this leads to a slightly lower median $\dot{M}_{\rm acc} \sim -7.7$ dex, suggesting that there is a non-negligible difference in detection limits of PMS stars at different $Z$.

This work is based on the assumption that the metallicity of the PMS stars in Sh2-284 is similar to that of the B stars located in the central open cluster Dolidze 25. The metallicity of Dolidze 25 is anomalous when taken into account the distance-metallicity gradient of the Galaxy (Rolleston et al. 2000). It has been suggested that the close association with the Canis Major Dwarf Galaxy is the reason for the metal-deficient nature of the central OB stars (Martin et al. 2004). It seems reasonable to assume that the metallicity of the PMS stars in the Sh2-284 region is similar to the central B stars, however, if other processes lead to the metal-deficient nature of the central B stars, the PMS stars may have higher abundances. Future work on the $Z$ determination of PMS stars in Sh2-284 would thus be welcomed.

## 5. SUMMARY AND FUTURE WORK

We presented an optical spectroscopic study of PMS stars located in the star-forming region Sh 2-284. Sh 2-284 is a Galactic star-forming region with $Z \sim 0.004$, i.e. 1/5 $Z_\odot$. Our sample consists of 24 objects spanning a mass range between 0.9 - 2.6 $M_\odot$ and an age range between 1 - 10 Myr, with a median age of 3.5 Myr. Our results are consistent with the scenario of sequential star formation suggested by Puga et al. (2009) inferred from gas dynamics. 21 stars have significant infrared excess up to 8$\mu$m, suggesting that they are bonafide Class I/II PMS stars, whilst the remaining PMS stars are most likely Class III objects.

We used the measured H$\alpha$ EW, and stellar parameters to derive $L_{\rm H\alpha}$, from which we estimated the $\dot{M}_{\rm acc}$. For five objects, we also have archival $U$-band photometry, and we measured the $U$-band excess intensity from which we estimated $\dot{M}_{\rm acc}$. We found that the $\dot{M}_{\rm acc}$ estimated from the $L_{\rm H\alpha}$ agree well with those derived from the $U$-band excess. The multiple signatures of disk accretion are highly suggestive of ongoing accretion in the Sh 2-284 PMS stars.

Our sample is detection limited in $L_{\rm acc}$ and $\dot{M}_{\rm acc}$. The highest measured $L_{\rm acc}$ is never greater than $L_*$, and the lowest measured $L_{\rm acc}$ corresponds to a H$\alpha$ EW $\sim -1$ Å. If accretion if present beyond such rates it is not easily detectable using current methods. We find $\dot{M}_{\rm acc}$ to steeply increase with $M_*$, with a power-law index $2.4 \pm 0.35$ when considering only stars in the 1 - 2 $M_\odot$ range. While $\dot{M}_{\rm acc}$ decrease with stellar age, with an index $-0.7 \pm 0.4$. There exists a spread $\sim 2$ orders of magnitude in the distribution of sources, which cannot be singularly attributed to any one effect.

Through comparison with solar-$Z$ PMS stars, we find no significant differences in the 1 - 2 $M_\odot$ range. The distribution of sources in the $\dot{M}_{\rm acc}$-$M_*$ and $\dot{M}_{\rm acc}$-$t_*$ planes are very similar, although there are fewer lower $\dot{M}_{\rm acc}$ objects at younger ages in Sh 2-284. However, this may be a product of low number statistics at very young ages. The median log $\dot{M}_{\rm acc}$ of Sh 2-284 ($-7.5$ dex) is in agreement with the median $\dot{M}_{\rm acc}$ of 3 Myr old solar-$Z$ star forming regions. When compared to the results in the LMC ($Z = 0.008$) from Spezzi et al. (2012), we find that the median log $\dot{M}_{\rm acc}$ of LMC PMS stars are $\sim -7.2$ dex at 6 Myr, suggesting that the median $\dot{M}_{\rm acc}$ are much lower in Sh 2-284. We also find that the distribution of Sh 2-284 sources in in the $\dot{M}_{\rm acc}$-$M_*$ and $\dot{M}_{\rm acc}$-$t_*$ planes occupy regions at lower $\dot{M}_{\rm acc}$ when compared to LMC stars. This difference is not easily explained, particularly when we consider that the $Z$ of the LMC is twice that of Sh 2-284, and if $Z$ is truly proportional to $\dot{M}_{\rm acc}$, $\dot{M}_{\rm acc}$ of Sh 2-284 PMS stars would be similarly higher than the LMC average, as the LMC measures are compared to the Galaxy. It is likely that the Spezzi et al. (2012) results are influenced by completeness issues. Moreover, it is not clear whether accretion can be sustained at such high rates for such long periods as measured in the LMC PMS stars, which, if true would suggest different disk evolutionary mechanisms in comparison to Galactic PMS stars, where accretion is thought to cease by around 5 Myr. We suggest that there are most likely factors beyond differences in $Z$ which influence the measured differences in $\dot{M}_{\rm acc}$. These include practical differences in detection limits at different $Z$, and difficulty in identifying PMS stars at distances to the LMC. A more systematic and thorough study of PMS stars in the Magellanic Clouds and the Galaxy is required to disentangle the various factors that influence measured mass-accretion rates.

Based on observations obtained with the South African Large Telescope under ID 2013-2-UKSC-008; and the ESO telescopes at the La Silla Paranal Observatory under program no. 0.74.C-0111. We would like to thank the referee and Drs. Antonella Natta and Danny Lennon for useful comments. V.M.K. is the recipient of a postgraduate scholarship awarded by the Department for Culture, Arts and Leisure Northern Ireland. J.S.V. and V.M.K. acknowledge funding from the the Science and Technology Facilities Council U.K.

## APPENDIX

### AGES AND $\dot{M}_{\rm ACC}$ DERIVED USING DIFFERENT STELLAR TRACKS AND ISOCHRONES

We use the PMS isochrones and tracks from Bressan et al. (2012) to derive the ages and masses for our sample. To account for differences in stellar models when comparing our results with literature results, we estimate the $\dot{M}_{\rm acc}$, $M_*$, and $t_*$ using the same Bressan et al. (2012) stellar tracks and isochrones at the appropriate $Z$. We used the quoted $\log L$, and $T_{\rm eff}$ of PMS stars in the LMC studies of Spezzi et al. (2012), and the Orion Nebula study by Manara et al. (2012) to estimate $M_*$ and $t_*$ using the Bressan et al. isochrones and tracks. The $L_{\rm acc}$ quoted by the Spezzi et al. (2012) and Manara et al. (2012) study were combined with the appropriate $M_*$ to derive $\dot{M}_{\rm acc}$ using Equation 2. In Figs. 9 and 10, we compare the $\dot{M}_{\rm acc}$ and ages derived using the Bressan et al. (2012) stellar models with those quoted in the studies. Namely, Spezzi et al. (2012) employed the Pisa isochrones database (Tognelli et al. 2011); while Manara et al. (2012) derived results using a variety of isochrones, where we chose the results they derived from employing the Palla & Stahler isochrones (1999). We find small systematic offsets in the ages and $\dot{M}_{\rm acc}$ when compared to the results of Spezzi et al. (2012), but no systematic differences in $\dot{M}_{\rm acc}$ compared to Manara et al. (2012). We find a non-systematic change in the ages of Orion Nebula

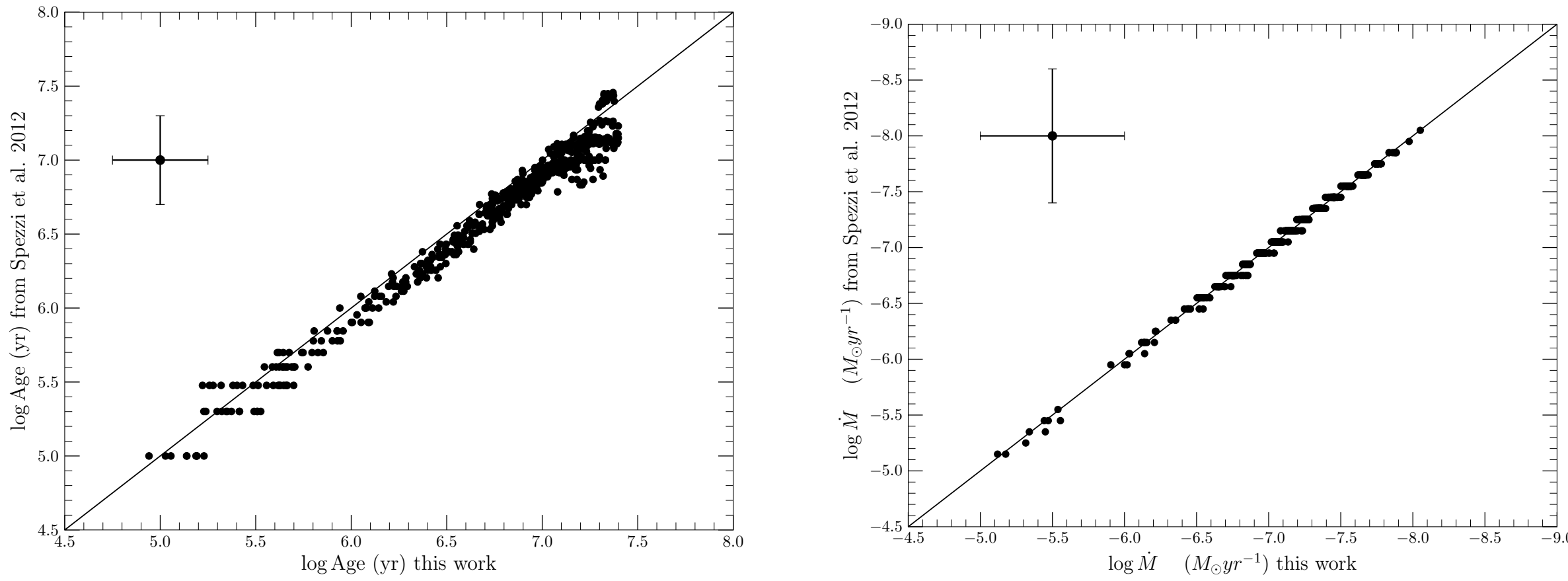

FIG. 9.— Comparison of accretion properties estimated using stellar parameters of PMS stars in the LMC (from Spezzi et al. 2012). Left: A comparison of the ages estimated by Spezzi et al. (2012) using Tognelli et al. (2011) isochrones (ordinate) and in this work using Bressan et al. (2012) isochrones (abscissa) at the appropriate $Z$. Right: A comparison of $\dot{M}_{\rm acc}$ derived using masses, radii using Bressan et al. (2012) isochrones (abscissa) versus those derived by Spezzi et al. (2012) (ordinate).

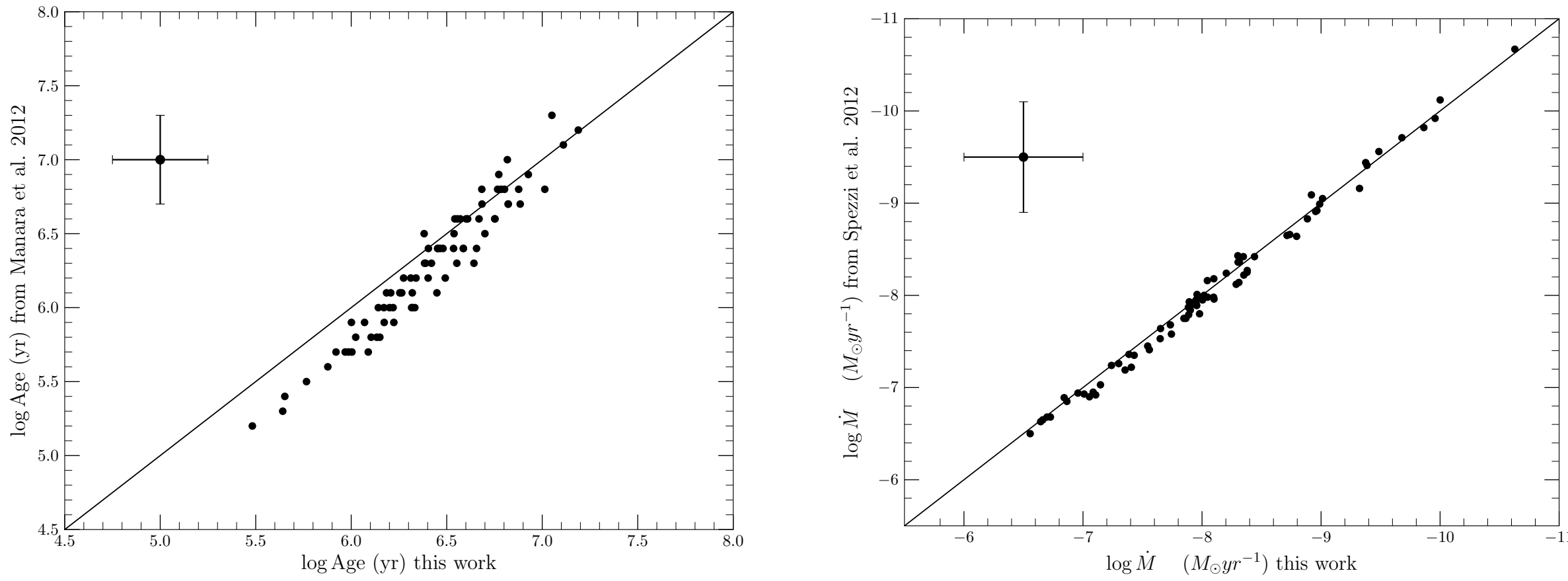

FIG. 10.— Comparison of accretion properties estimated using stellar parameters of PMS stars in the Orion Nebula (from Manara et al. 2012). Left: A comparison of the ages estimated by Manara et al. (2012) using Palla & Stahler (1999) isochrones (ordinate) and in this work using Bressan et al. (2012) isochrones (abscissa). Right: A comparison of $\dot{M}_{\rm acc}$ derived using masses, radii from Bressan et al. (2012) isochrones, and $L_{\rm H\alpha}$-$L_{\rm acc}$ relation used in this work (abscissa) versus those derived by Manara et al. (2012) (ordinate). The $L_{\rm acc}$ used to derive the $\dot{M}_{\rm acc}$ was taken from Manara et al. (2012).

PMS stars less than 1 Myr using different isochrones.

## TESTING THE $\dot{M}_{\rm ACC}$ ESTIMATION METHOD

We demonstrate that the method used to calculate $\dot{M}_{\rm acc}$ reproduce the $\dot{M}_{\rm acc}$ estimated from a direct measurement of the H$\alpha$ line flux in Fig. 11. We used the H$\alpha$ EW and stellar parameters ($T_{\rm eff}$, $M_*$, $R_*$, $A_V$, $d$) of eight Herbig Ae/Be stars determined by Pogodin et al. (2012), to calculate their $\dot{M}_{\rm acc}$ using the method described in Section 3.2 and 3.3. We compare the $\dot{M}_{\rm acc}$ calculated this way, with those estimated from the intensity of H$\alpha$ emission line measured from flux-calibrated spectra by Pogodin et al. (2012). We find that our results accurately reproduce the measured $\dot{M}_{\rm acc}$ within 0.15 dex, the error due to uncertainties in stellar parameters and measured H$\alpha$ EW.

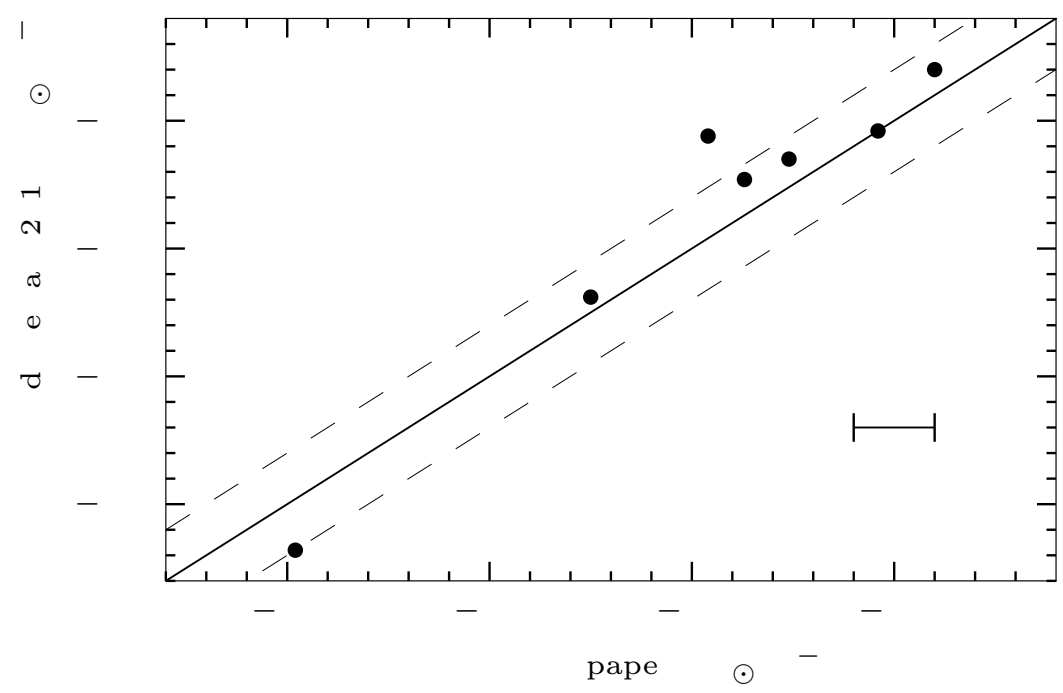

FIG. 11.— Comparison of $\dot{M}_{\rm acc}$ (abscissa) derived using the method followed in this paper versus traditional line diagnostic methods (ordinate). All data taken from Pogodin et al. (2012).